\documentclass[conference,9pt]{IEEEtran}

\IEEEoverridecommandlockouts

\usepackage{cite}
\usepackage{amsmath,amssymb,amsfonts}
\usepackage{algorithmic}
\usepackage{graphicx}
\usepackage{textcomp}
\usepackage[dvipsnames]{xcolor}
\usepackage{paralist}
\usepackage[inline]{enumitem}
\usepackage{caption}
\usepackage{multirow}
\usepackage{booktabs}
\usepackage{makecell}
\usepackage{tabu, booktabs}
\usepackage{float}
\usepackage{subcaption}
\usepackage{pifont}
\usepackage{comment}
\usepackage{todonotes}
\usepackage{hyperref}
\usepackage{csquotes}

\newcommand\myshade{85}

\definecolor{magma_darker}{HTML}{fdc38a}
\definecolor{magma_dark}{HTML}{e15666}
\definecolor{magma_lighter}{HTML}{1f0c43}

\definecolor{cvpr_link}{HTML}{00ffff}
\definecolor{cvpr_cite}{HTML}{00ff00}
\definecolor{cvpr_file}{HTML}{ff0000}

\definecolor{royalblue}{RGB}{65, 105, 225}
\definecolor{darkgreen}{RGB}{0, 100, 0}
\definecolor{darkred}{RGB}{139, 0, 0}
\definecolor{darkviolet}{RGB}{148, 0, 211}

\colorlet{linkColor}{violet}
\colorlet{citeColor}{YellowOrange}
\colorlet{urlColor}{Emerald}

\hypersetup{
  linkcolor  = darkviolet!\myshade!black,
  citecolor  = citeColor!\myshade!black,
  urlcolor   = urlColor!\myshade!black,
  filecolor  = darkgreen!\myshade!black,
  colorlinks = true,
}
\usepackage[a4paper, total={184mm,239mm}]{geometry}
\def\BibTeX{{\rm B\kern-.05em{\sc i\kern-.025em b}\kern-.08em
    T\kern-.1667em\lower.7ex\hbox{E}\kern-.125emX}}

\ExplSyntaxOn
\DeclareExpandableDocumentCommand{\convertlen}{ O{cm} m }
 {
  \dim_to_decimal_in_unit:nn { #2 } { 1 #1 } cm
 }
\ExplSyntaxOff
\usepackage{gnuplottex}
\usepackage{tikz}
\usepackage{mathtools}

\usetikzlibrary{matrix,calc,positioning,arrows.meta}

\newcommand*\circled[1]{\tikz[baseline=(char.base)]{\node[shape=circle,fill,inner sep=0.5pt] (char) {\textcolor{white}{#1}};}}

\begin{document}

\title{Carbon-Efficient 3D DNN Acceleration: Optimizing Performance and Sustainability}
\author{
\IEEEauthorblockN{
Aikaterini Maria Panteleaki\IEEEauthorrefmark{1},
Konstantinos Balaskas\IEEEauthorrefmark{2},
Georgios Zervakis\IEEEauthorrefmark{2},
Hussam Amrouch\IEEEauthorrefmark{3},
Iraklis Anagnostopoulos\IEEEauthorrefmark{1}
}
\IEEEauthorblockA{
\IEEEauthorrefmark{1}Southern Illinois University Carbondale, 
\IEEEauthorrefmark{2}University of Patras, 
\IEEEauthorrefmark{3}Technical University of Munich
}
}

\maketitle

\begin{abstract}
As Deep Neural Networks (DNNs) continue to drive advancements in artificial intelligence, the design of hardware accelerators faces growing concerns over embodied carbon footprint due to complex fabrication processes. 3D integration improves performance but introduces sustainability challenges, making carbon-aware optimization essential. In this work, we propose a carbon-efficient design methodology for 3D DNN accelerators, leveraging approximate computing and genetic algorithm-based design space exploration to optimize Carbon Delay Product (CDP). By integrating area-efficient approximate multipliers into Multiply-Accumulate (MAC) units, our approach effectively reduces silicon area and fabrication overhead while maintaining high computational accuracy. Experimental evaluations across three technology nodes (45nm, 14nm, and 7nm) show that our method reduces embodied carbon by up to 30\% with negligible accuracy drop. 
\end{abstract}

\begin{IEEEkeywords}
DNN Accelerators, 3D Integration, Approximate Computing, Embodied Carbon Footprint, Sustainable Computing
\end{IEEEkeywords}

\setlist{nosep}

\section{Introduction}\label{sec:intro}

The rapid growth of Artificial Intelligence (AI) has resulted in the wide adoption of Deep Neural Networks (DNNs) as a fundamental component of modern computing systems. To efficiently support the computational demands of DNNs, specialized hardware accelerators have been developed, offering significant improvements in throughput and energy efficiency. These accelerators have enabled AI deployment across a wide range of environments, from large-scale data centers to resource-constrained edge devices. However, as AI applications continue to scale in complexity, the environmental impact of DNN accelerators has become a growing concern ~\cite{gupta2021chasing}. While prior optimization efforts have focused on reducing operational energy consumption, recent studies indicate that embodied carbon, arising from semiconductor manufacturing, packaging, and fabrication processes, can contribute significantly to the total carbon footprint of AI hardware, particularly for edge devices ~\cite{panteleaki2024carbon,gupta2022act}. Addressing these sustainability challenges requires novel architectural strategies that balance performance with carbon efficiency.

To enhance performance and energy efficiency, many hardware designers have adopted three-dimensional (3D) integration as a promising architectural strategy for DNN accelerators. Unlike standard two-dimensional (2D) designs, 3D integration enables vertical stacking of multiple chiplets using advanced packaging techniques such as hybrid bonding and Through-Silicon Vias (TSVs)~\cite{jeloka2022system}. This vertical integration reduces interconnect lengths, increases memory bandwidth, and minimizes chip footprint, making it particularly beneficial for resource-constrained environments like edge and mobile computing systems. By leveraging 3D architectures, accelerators can integrate more computational and memory resources within the same physical footprint, achieving lower latency and reduced power consumption compared to their 2D counterparts~\cite{yang2022three}. However, despite these performance advantages, the manufacturing complexity of 3D integration introduces additional sustainability challenges. The embodied carbon footprint of 3D accelerators is significantly higher than that of equivalent 2D designs due to increased wafer processing steps, bonding requirements, and lower fabrication yields~\cite{byun2024energy}. As a result, there exists a critical trade-off between adopting 3D architectures for their computational advantages and mitigating their environmental impact.

To address this sustainability challenge, recent methods have explored the trade-offs between performance and carbon footprint in DNN accelerator design, primarily focusing on optimizing 2D architectures~\cite{panteleaki2024carbon,panteleaki2025late}. While these studies provide valuable insights, the transition to 3D architectures creates embodied carbon concerns due to additional manufacturing overhead. Specifically, processes such as wafer thinning, TSV etching, and bonding not only increase energy consumption during fabrication but also contribute to material waste and reduced yields~\cite{jeloka2022system}. This issue is particularly important in smaller accelerators designed for edge applications, where the benefits of reduced chip area may be outweighed by the higher carbon cost of 3D integration~\cite{byun2024energy}. As AI accelerators support more and more mobile and embedded platforms, finding sustainable design methodologies for 3D accelerators becomes essential to balance computational efficiency with embodied carbon footprint.

Approximate computing has been widely explored as a technique for reducing energy consumption in DNN accelerators, leveraging the inherent error resilience of deep learning models~\cite{tasoulas2020weight}. Since minor inaccuracies in arithmetic computations typically have minimal impact on inference accuracy, approximate multipliers have been introduced in Multiply-Accumulate (MAC) units to achieve power and energy efficiency~\cite{spantidi2021positive}. However, despite its success in lowering operational energy, approximate computing has not yet been explored as a strategy to mitigate embodied carbon. By replacing exact multipliers in MAC units with carefully selected approximate counterparts, it is possible to reduce silicon area, fabrication complexity, and material usage, all of which contribute directly to the embodied carbon footprint. Unlike conventional energy-focused optimizations, this approach explicitly targets the environmental cost of semiconductor manufacturing, making it particularly beneficial for 3D-integrated architectures, where embodied carbon is a dominant factor.

In this work, we present a carbon-aware design methodology for 3D DNN accelerators that leverages approximate computing to reduce embodied emissions while maintaining high computational efficiency. We explore a 3D memory-on-logic architecture, where the bottom die houses computational logic, including a systolic array of Processing Elements (PEs), while the top die integrates a global SRAM buffer for storing input activations, weights, and output feature maps\cite{yang2022three,wu202411}. Since MAC units are among the most area-intensive components, we replace conventional multipliers with area-efficient approximate multipliers, significantly reducing silicon usage while maintaining acceptable accuracy for DNN inference. To systematically optimize this trade-off between performance, accuracy, and sustainability, we employ a genetic algorithm-based design space exploration framework, which searches for architectures that minimize Carbon Delay Product (CDP) while ensuring efficient computation. The key contributions of this paper are as follows:
\begin{inparaenum}
    \item[\circled{1}] We introduce a sustainability-driven approximation strategy that replaces exact multipliers in MAC units with optimized approximate counterparts, reducing silicon area and the associated embodied carbon footprint.
    \item[\circled{2}] We develop a multi-objective optimization framework that explores key architectural parameters, such as the number of Processing Elements (PEs), local buffer sizes, and memory hierarchy—while balancing performance, accuracy, and sustainability metrics.
    \item[\circled{3}] We define and optimize CDP as a metric that simultaneously considers both embodied carbon footprint and computational efficiency, ensuring that the resulting accelerator designs achieve a sustainable balance between performance and embodied carbon emissions.
\end{inparaenum}
\section{Related Work}

The design space exploration of DNN accelerators has traditionally focused on optimizing timing, power, and area. To improve efficiency, prior works have explored co-optimization of hardware architecture and DNN mapping strategies. In~\cite{kao2022digamma}, a genetic algorithm-based framework optimizes hardware parameters and mapping for 2D DNN accelerators. Alternatively,\cite{hong2023dosa} employs gradient descent to navigate the unified design space. Moving beyond monolithic 2D accelerators,\cite{gao2019tangram} introduces a tiled accelerator paradigm, enabling both inter-layer and layer-by-layer optimizations. With the rise of 3D integration, design space exploration has expanded to multi-chiplet architectures. In~\cite{mishty2024chiplet}, both Reinforcement Learning (RL) and non-RL algorithms optimize chiplet-based AI accelerators for performance, power, area, and cost, targeting Large Language Models (LLMs). 
The evaluation framework in\cite{byun20243d} analyzes 2D and 3D systolic arrays, emphasizing energy consumption, while~\cite{byun2024energy} extends this work by integrating sustainability analysis, assessing both operational and embodied carbon emissions.


For a holistic assessment of DNN accelerator designs, many researches have highlighted the importance of sustainable hardware approaches. The work in \cite{gupta2022act} introduces an architectural carbon footprint modeling framework that enables early-stage design space exploration. Building on this foundation, \cite{elgamal2023design} investigates carbon-aware design parameters for virtual and extended reality systems, while jointly optimizing operational and embodied carbon alongside with job execution latency.  However, recent studies \cite{bashir2024promise} have demonstrated that embodied and operational emissions are estimated on different scales and therefore cannot be directly compared.
Focusing specifically on DNN accelerators, \cite{panteleaki2024carbon} explores designs that balance the critical trade-off between sustainability and performance for 2D systolic array architectures. More advanced modeling tools presented in \cite{zhao20243d} and \cite{sudarshan2024eco} provide detailed carbon assessments for 2.5D and 3D chips, incorporating crucial manufacturing parameters, such as technology node,  yield, packaging type and most importantly the chip area. While these works provide valuable carbon modeling methodologies, our approach proposes approximate computing as a strategy to mitigate embodied carbon footprint of 3D DNN accelerators, offering a novel approach to design sustainability-aware devices with high performance.

Approximate computing leverages the inherent error tolerance of deep neural networks, as computational imprecisions in operations can have minimal impact on the overall accuracy. In the context of DNN accelerators, a wide variety of works demonstrate the benefits of approximate computing, leading to significant reductions in energy consumption, job delay and area footprint, while maintaining acceptable accuracy levels. The works in \cite{spantidi2021positive}, \cite{spantidi2023perfect} explore the trade-offs between energy efficiency and accuracy loss when incorporating reconfigurable approximate multipliers in the accelerator logic blocks. The authors in \cite{10924159} investigate the accuracy impact of approximate circuits for various types of Convolutional Neural Networks, while the tools presented in \cite{vaverka2020tfapprox,gong2023approxtrain} offer complete frameworks that evaluate the training or inference accuracy of a DNN model when incorporating approximate multipliers. While these works highlight the gains from using approximate circuits in DNN applications, they examine only 2D accelerator architectures. Our work uniquely applies approximation in 3D DNN accelerators, as a strategy to reduce the embodied carbon footprint and address the sustainability challenges that are very common in 3D integrated systems.
\section{Methodology}\label{sec:methodology}

\begin{figure}
    \centering
    \resizebox{0.45\textwidth}{!}{\includegraphics[width=\linewidth, clip]{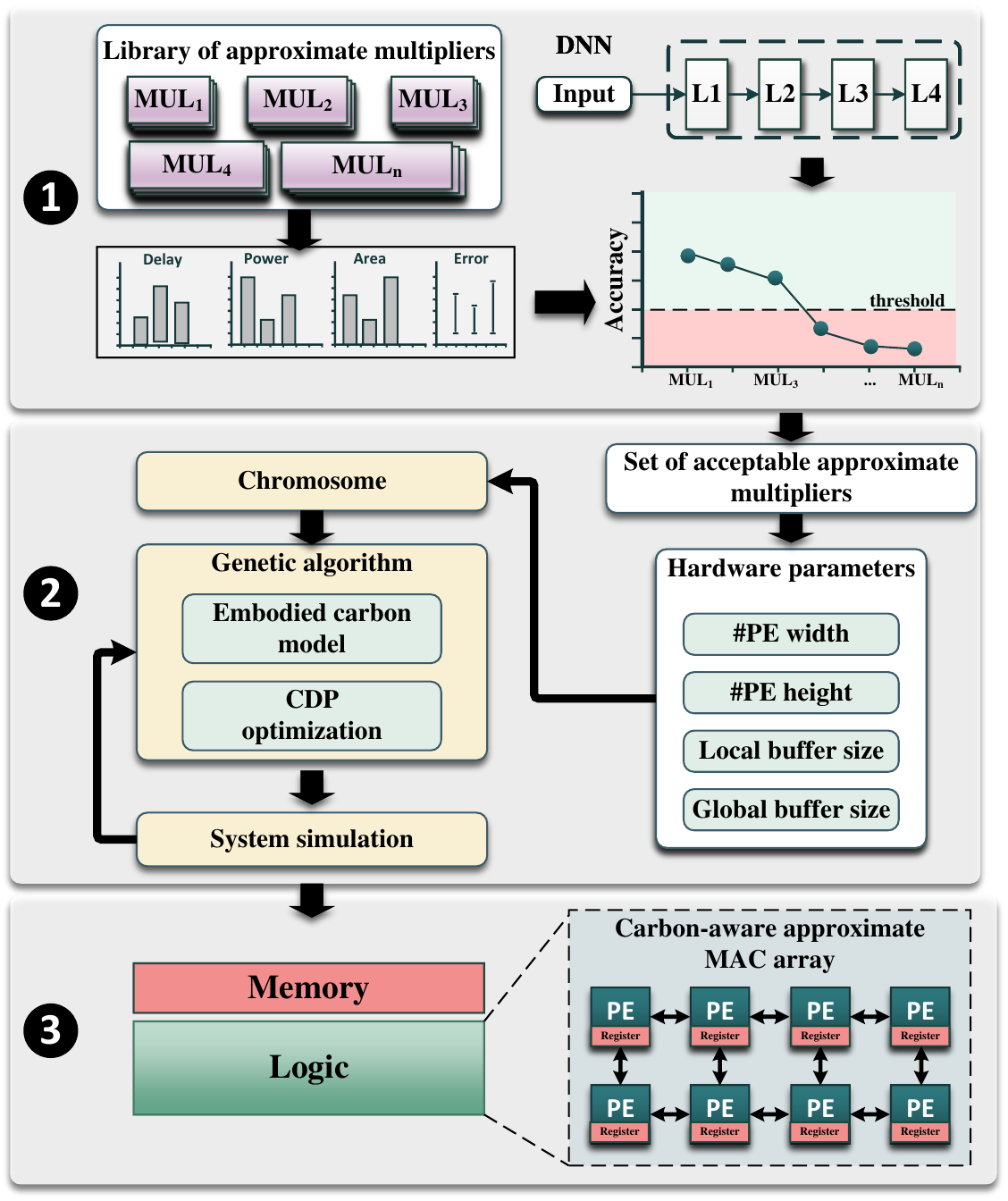}}
    \caption{Overview of the proposed methodology}
    \vspace{-10pt}
    \label{fig:overview}
\end{figure}

Figure~\ref{fig:overview} presents an overview of our proposed methodology for designing carbon-aware 3D DNN accelerators using approximate computing and a genetic algorithm-based optimization strategy. The first stage involves a library of approximate multipliers, where various multiplier designs are evaluated based on delay, power, area, and error, ensuring that only configurations meeting predefined accuracy thresholds are selected. The second stage integrates these approximate multipliers into the hardware design space exploration, where a genetic algorithm optimizes the Carbon Delay Product (CDP) by selecting the best combination of Processing Element (PE) configurations, buffer sizes, and approximate multipliers. Finally, the optimized carbon-aware 3D accelerator is evaluated, to verify the gains in embodied carbon without performance compromises.

\subsection{3D memory-on-logic accelerator}\label{sec:3d_design}

For our framework, we adopt a DNN accelerator model based on the Eyeriss architecture \cite{chen2016eyeriss}, which features a  mesh-based Processing Element (PE) array  and a  hierarchical memory system  optimized for deep neural network (DNN) workloads. Each PE includes a  Multiply-Accumulate (MAC) unit  responsible for performing core computations, as well as a  local buffer  in the form of a register file that stores DNN weights and intermediate results. To facilitate efficient data movement, the accelerator exchanges data with  off-chip DRAM  through a  global SRAM memory , which is accessible by all PEs and is used to store input activations, weights, and output feature maps.

In conventional 2D accelerator designs, data transfer between the global SRAM and the PEs relies on a Network-on-Chip (NoC), introducing communication overhead and energy inefficiencies. However, our design extends this architecture by leveraging memory-on-logic 3D integration paradigm, where the logic layer (bottom die) hosts the PE array and compute elements, while the memory layer (top die) integrates the global SRAM. This 3D stacking allows for tighter coupling between processing and memory resources, reducing data transfer latency and energy consumption.

For vertical connectivity between the two dies, we utilize   hybrid bonding technology, a state-of-the-art 3D integration technique that provides   higher bandwidth   and   lower interconnect resistance   compared to traditional methods like microbumps~\cite{wu202411}. This approach enhances performance by enabling   denser and more energy-efficient inter-die communication. Additionally, the final   3D packaging   employs   Through-Silicon-Vias (TSVs), which establish   direct electrical connections   between the stacked dies and the package substrate. These TSVs further optimize   data movement efficiency, which is particularly beneficial for   memory-intensive DNN workloads  . By minimizing   off-chip memory accesses, our design significantly improves   performance and energy efficiency, making it well-suited for sustainable DNN acceleration.

\subsection{Embodied Carbon in 3D chips}\label{sec:embodied_carbon}

The \textit{embodied carbon footprint} represents the total $\text{CO}_2$ emissions associated with a device over its entire life cycle. However, for semiconductor-based systems, it is primarily dominated by emissions from fabrication processes in semiconductor manufacturing facilities~\cite{gupta2022act}. In this work, we analyze the embodied carbon footprint of 3D DNN accelerators by considering emissions from the fabrication of the logic die, memory die, as well as the bonding and packaging processes that occur in fabrication plants. To quantify the embodied carbon emissions, we follow the analytical models proposed in~\cite{byun20243d, sudarshan2024eco}, which provide fine-grained estimations for 3D-integrated circuits. The total embodied carbon of a 3D DNN accelerator is formulated as:
\begin{equation}
    C_{\text{embodied}} = C_{\text{die}}^{\text{logic}} + C_{\text{die}}^{\text{memory}} + C_{\text{bonding}} + C_{\text{packaging}} 
    \label{eq:emb}
\end{equation}
where:
\begin{inparaenum}[(i)]
    \item $C_{\text{die}}^{\text{logic}}$ and $C_{\text{die}}^{\text{memory}}$ represent the carbon emissions from the fabrication of the logic and memory dies accordingly;
    \item $C_{\text{bonding}}$ represents the emissions from 3D bonding techniques (e.g., hybrid bonding, microbumps); and
    \item $C_{\text{packaging}}$ represents the emissions from final chip packaging and interconnect integration.
\end{inparaenum}
For each die, the fabrication-related carbon emissions are calculated as:
\begin{equation}
    C_{\text{die}} = \text{CFPA} \times A_{\text{die}} + \text{CFPA}_{\text{Si}} \times A_{\text{wasted}}
    \label{eq:carbon_die}
\end{equation}
where:
\begin{inparaenum}[(i)]
    \item $\text{CFPA}$ represents the carbon footprint per unit area of the die;
    \item $A_{\text{die}}$ is the effective area of the manufactured chip;
    \item $\text{CFPA}_{\text{Si}}$ accounts for silicon wastage due to the wafer dicing process; and
    \item $A_{\text{wasted}}$ is the unused silicon area on the wafer.
\end{inparaenum}
The carbon footprint per unit area (CFPA) is defined as:
\begin{equation}
    \text{CFPA} =  \frac{\text{CI}_\text{fab} \times \text{EPA} + \text{C}_\text{gas} + \text{C}_\text{material}}{\text{Y}}
    \label{eq:cfpa}
\end{equation}
where:
\begin{inparaenum}[(i)]
    \item $\text{CI}_\text{fab}$ is the carbon intensity of the fabrication facility (depends on the energy sources used in the fab);
    \item $\text{EPA}$ is the energy per unit area required for manufacturing;
    \item $\text{C}_\text{gas}$ represents the greenhouse gas emissions from fabrication;
    \item $\text{C}_\text{material}$ is the carbon cost of raw material procurement; and
    \item $\text{Y}$ represents the fraction of successfully fabricated dies (yield percentage).
\end{inparaenum}
Similarly, the embodied carbon associated with 3D bonding and packaging is proportional to the respective surface areas involved:
\begin{equation}
    C_{\text{bonding}} =  \text{CFPA}_{\text{bonding}} \times A_{\text{die}}
    \label{eq:bonding}
\end{equation}
\begin{equation}
    C_{\text{packaging}} =  \text{CFPA}_{\text{packaging}} \times A_{\text{package}}
    \label{eq:packaging}
\end{equation}
where $\text{CFPA}_{\text{bonding}}$ and $\text{CFPA}_{\text{packaging}}$ define the carbon cost per unit area for bonding and packaging processes, respectively.

\subsection{Die Area Estimation for Embodied Carbon Modeling}

From the above equations, it is clear that the chip area is the dominant factor affecting the embodied carbon emissions of a 3D accelerator. For the memory die, which consists of the global SRAM buffer, we utilize CACTI~\cite{CACTI} to estimate its area based on memory capacity and technology node. In cases where CACTI does not support a specific target technology node, we apply area scaling trends from~\cite{sudarshan2024eco} to extrapolate the required values.

The logic die area, on the other hand, is determined primarily by the number of PEs in the design. Each PE contains a local buffer and a Multiply-Accumulate (MAC) unit, both of which contribute significantly to the total silicon footprint. The local buffer, implemented as a register file, follows a similar area estimation approach to that of the global SRAM, relying on existing area scaling models. However, the MAC unit is the most area-intensive component of the PE and thus has a significant impact on the overall logic die area.

To further break down the MAC unit's area composition, we analyze its arithmetic circuits under the widely used bfloat16 representation, which has been extensively adopted in modern DNN accelerators, such as Google's TPUs~\cite{bfloat16}. Each MAC unit consists of several key arithmetic blocks, including a 7-bit multiplier for mantissa multiplication, two 8-bit adders for exponent addition, and a 24-bit integer adder (23 bits mantissa plus the hidden bit) for accumulating the mantissa results. Among these components, multipliers are by far the most silicon-intensive, dominating the area requirements of the MAC unit. Given their substantial contribution to the overall die size, reducing the area of multipliers directly impacts the total embodied carbon footprint of the accelerator. This observation motivates our focus on employing approximate multipliers, as their adoption can lead to significant reductions in silicon area, thereby lowering the accelerator's carbon footprint while maintaining computational efficiency.

\subsection{Approximate multipliers in 3D MAC units}\label{sec:approx_mac}

To mitigate the embodied carbon footprint of 3D DNN accelerators, we employ \textit{approximate computing} as a primary strategy to reduce the silicon area of the circuits that form the Multiply-Accumulate (MAC) units. Among the arithmetic components in a MAC unit, multipliers are the most resource-intensive in terms of both area and power consumption. Unlike adders, multipliers require a significantly larger number of logic gates, making them ideal candidates for approximation. Importantly, deep neural networks (DNNs) exhibit a high tolerance to approximation due to their inherent redundancy and statistical properties. The error resilience of DNN workloads arises from two key factors. First, the inference process relies more on relative weight distributions rather than precise numerical values, meaning that small deviations in computation often do not substantially alter the model's final output~\cite{tasoulas2020weight}. Second, the accumulation of multiple operations in deep architectures leads to an averaging effect, where individual approximation errors tend to cancel out rather than propagate destructively~\cite{spantidi2021positive}.

In our design, we replace conventional exact multipliers with area-efficient \textit{approximate multipliers} from the EvoApprox library~\cite{mrazek2017evoapprox8b}, which provides a wide range of arithmetic circuits with different trade-offs between computational accuracy and area. Specifically, we target the \textit{7-bit multipliers} responsible for \textit{mantissa multiplication} in the bfloat16 MAC unit. By employing approximate versions of these multipliers, we significantly reduce the logic complexity, leading to lower transistor count and smaller die area. Given that the embodied carbon footprint of a die is proportional to its silicon area, this reduction translates directly into lower $\text{CO}_2$ emissions. To ensure that the use of approximate multipliers does not significantly lower  model accuracy, we utilize \textit{ApproxTrain}~\cite{gong2023approxtrain}, a framework designed to systematically simulate the propagation of approximation errors in DNN architectures. ApproxTrain evaluates the impact of replacing exact arithmetic operations with approximate versions, allowing us to quantify the trade-offs between accuracy and area savings.

\subsection{Genetic algorithm-based design space exploration}\label{sec:ga}

To systematically explore the design space of our 3D accelerator and guide the carbon optimization process, we develop a \textit{multi-objective genetic algorithm} that efficiently navigates the vast configuration space to find optimal solutions that minimize the \textit{Carbon Delay Product (CDP)}. This approach allows us to select accelerator configurations that effectively balance the trade-off between performance and embodied carbon footprint, a particularly challenging problem in 3D architectures due to increased fabrication complexity and energy costs. The genetic algorithm represents each accelerator configuration as a \textit{chromosome}, encoding key hardware parameters, including the Processing Element (PE) array dimensions, local buffer size, and global SRAM capacity. Each chromosome is structured as:
\begin{equation}
    \mathbf{C} = \{ P_x, P_y, B_{\text{local}}, B_{\text{global}} \}
\end{equation}
where:
\begin{inparaenum}[(i)]
    \item $P_x, P_y$ denote the dimensions of the PE array;
    \item $B_{\text{local}}$ represents the local buffer size per PE; and
    \item $B_{\text{global}}$ represents the global SRAM capacity.
\end{inparaenum}

Our framework integrates nn-dataflow~\cite{gao2019tangram} for mapping exploration, performing delay-optimized dataflow scheduling for specified hardware architectures and DNN models. We extended nn-dataflow to support memory-on-logic 3D architectures, which leverage vertical high-bandwidth interconnects between the PE array and the global SRAM. By incorporating these low-latency connections, the tool accurately models the reduced communication cost between the logic and memory layers, optimizing the dataflow strategy accordingly. 

As aforementioned, our framework integrates EvoApprox approximate multipliers~\cite{mrazek2017evoapprox8b}, selecting the most area-efficient multiplier that satisfies an accuracy drop constraint. For each target DNN, we impose a maximum allowable accuracy degradation of 1\%, 2\%, and 3\%, ensuring that the accelerator maintains acceptable inference performance. Given a neural network and its associated accuracy constraint $\delta$, the framework selects the most area-efficient approximate multiplier $M_{\text{approx}}$ that satisfies:
\begin{equation}
    \Delta A(M_{\text{approx}}) \leq \delta
    \label{eq:accuracy_constraint}
\end{equation}
where $\Delta A(M_{\text{approx}})$ represents the accuracy drop introduced by using a specific approximate multiplier $M_{\text{approx}}$ compared to an exact multiplier.


The genetic algorithm follows a structured evolutionary process to explore the accelerator design space. Before the optimization process, \textit{ApproxTrain simulations}~\cite{gong2023approxtrain} are used to evaluate the accuracy impact of different approximate multipliers. Only multipliers that satisfy the predefined accuracy constraints (Eq.~\ref{eq:accuracy_constraint}) are included in the design space. In \textit{Step 1: Initialization}, a population of $N$ candidate configurations (chromosomes) is randomly generated. In \textit{Step 2: Fitness Evaluation}, each candidate is assessed based on multiple criteria: the \textit{carbon model} (Eq.~\ref{eq:emb}) is used to compute $C_{\text{embodied}}$, and the \textit{nn-dataflow performance model} estimates $D_{\text{task}}$. In \textit{Step 3: Selection}, the top-performing designs with the lowest CDP values are chosen for reproduction. In \textit{Step 4: Crossover}, new candidate configurations are generated by recombining parameter values from selected solutions. To introduce diversity and prevent premature convergence, \textit{Step 5: Mutation} applies random modifications to some parameters. Finally, in \textit{Step 6: Termination}, the algorithm iterates for $G$ generations until convergence criteria are met, ensuring that the best accelerator designs are identified based on both computational efficiency and embodied carbon reduction.

\section{Evaluation}\label{sec:evaluation}

We evaluated our framework considering inference delay  and embodied carbon footprint across three technology nodes: 45nm, 14nm, and 7nm. These nodes were selected to show scalability and applicability to modern fabrication processes. The 45nm node, used in the EvoApprox library~\cite{mrazek2017evoapprox8b}, provides post-synthesis area data for approximate multipliers. For 14nm and 7nm, we synthesized the approximate multipliers using the Synopsys Design Compiler to obtain accurate area estimations. The clock frequencies for each node are scaled appropriately with existing technology advancements and set to 500 MHz (45nm), 940 MHz (14nm), and 1050 MHz (7nm).

To evaluate the impact of the approximate multipliers on DNN inference accuracy, we integrated their netlists into ApproxTrain and tested them on ImageNet dataset. We selected five widely used CNNs, VGG16, VGG19, ResNet50, ResNet50V2, and DenseNet, due to their architectural diversity and utilization in edge computing applications, particularly in AR/VR workloads where resource efficiency is critical. To ensure a balance between accuracy and carbon reduction, we considered three accuracy drop thresholds: 1\%, 2\%, and 3\%, selecting only the approximate multipliers that maintained accuracy within these limits each time. 

For embodied carbon estimation, we utilized detailed manufacturing parameters from~\cite{wu202411}, which models a 3D DNN System-on-Chip incorporating hybrid bonding for inter-chiplet connections, TSV packaging, and Face-to-Face wafer stacking. The Carbon Footprint Per Area (CFPA), area scaling trends, and yield values for different technology nodes are derived from the methodologies presented in~\cite{zhao20243d, sudarshan2024eco}, providing a comprehensive foundation for accurate carbon footprint assessment across the different selected fabrication processes.

\subsection{Delay and Embodied Carbon comparison}

\begin{figure*}[ht!]
\centering
  \begin{subfigure}{0.49\linewidth}
    \centering
    \includegraphics[width=\linewidth]{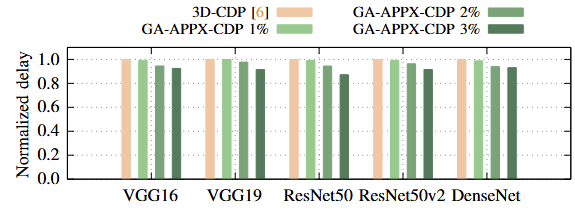}
    \vspace*{-1.5em}
    \caption{Normalized Delay (45 nm)}
    \label{fig:45nm-delay}
  \end{subfigure}
  \begin{subfigure}{0.49\linewidth}
    \centering
    \includegraphics[width=\linewidth]{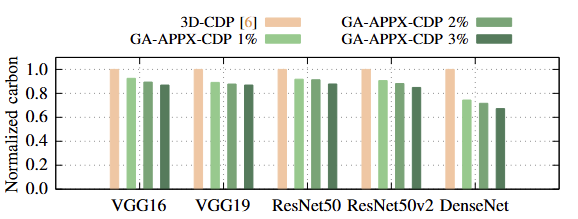}
    \vspace*{-1.5em}
    \caption{Normalized Embodied Carbon (45 nm)}
    \label{fig:45nm-carbon}
  \end{subfigure}
  
  \begin{subfigure}{0.49\linewidth}
    \centering
    \includegraphics[width=\linewidth]{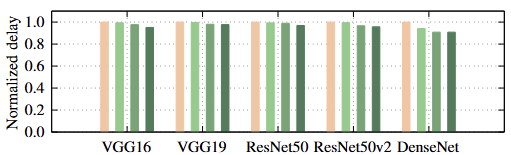}
    \vspace*{-1.5em}
    \caption{Normalized Delay (14 nm)}
    \label{fig:14nm-delay}
  \end{subfigure}
  \begin{subfigure}{0.49\linewidth}
    \centering
    \includegraphics[width=\linewidth]{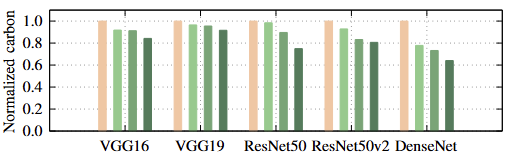}
    \vspace*{-1.5em}
    \caption{Normalized Embodied Carbon (14 nm)}
    \label{fig:14nm-carbon}
  \end{subfigure}

  \begin{subfigure}{0.49\linewidth}
    \centering
    \includegraphics[width=\linewidth]{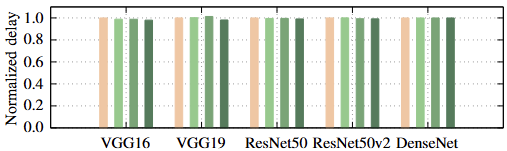}
    \vspace*{-1.5em}
    \caption{Normalized Delay (7 nm)}
    \label{fig:7nm-delay}
  \end{subfigure}
  \begin{subfigure}{0.49\linewidth}
    \centering
    \includegraphics[width=\linewidth]{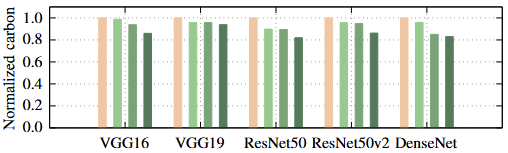}
    \vspace*{-1.5em}
    \caption{Normalized Embodied Carbon (7 nm)}
    \label{fig:7nm-carbon}
  \end{subfigure}
  
  \caption{Comparison of normalized inference delay and embodied carbon across different technology nodes (45nm, 14nm, and 7nm). GA-APPX-CDP is our proposed method for different accuracy drop thresholds.}
  \label{fig:tech-node-comparison}
\end{figure*}

In Figure~\ref{fig:tech-node-comparison}, we show the results in terms of normalized delay and normalized carbon for all three selected technology nodes (45nm, 14nm, and 7nm) and three accuracy drop thresholds (1\%, 2\%, and 3\%). As a baseline, we used the optimization approach presented in~\cite{byun2024energy}, which explores CDP optimization for 3D DNN accelerators without utilizing approximate computing techniques. Our results demonstrate that the integration of approximate multipliers along with the genetic algorithm-based design space exploration
consistently reduces the embodied carbon footprint across all configurations while maintaining competitive performance.

The most significant reductions are observed at 45nm and 14nm, where our framework, \emph{labeled as GA-APPX-CDP}, achieves up to 25\% and 30\% lower embodied carbon, respectively, compared to the baseline. At 7nm, the improvements are more limited, reaching only 15\%, as the relative impact of silicon area reductions diminishes in smaller process nodes due to increased interconnect overhead and static power dissipation. The benefits of approximate computing are particularly evident in compute-bound models, such as DenseNet, where carbon reductions reach 30\%, driven by a fundamental shift in the design space exploration process. By reducing MAC unit area, the genetic algorithm favors configurations with higher PE counts, increasing parallelism and reducing inference latency while maintaining efficient workload distribution. This results in a positive feedback loop, where additional PEs not only accelerate computations but also improve sustainability by distributing processing more efficiently. In contrast, exact computing baselines require larger, more carbon-intensive units, limiting their ability to optimize both performance and carbon footprint simultaneously. The relationship between accuracy drop thresholds (1\%, 2\%, and 3\%) and embodied carbon follows a clear trend, where higher tolerance for approximation leads to greater area savings, directly reducing the carbon footprint. These results show that approximate computing enables a co-optimization of computational efficiency and sustainability, making it a highly effective strategy for 3D-integrated DNN accelerators.

\subsection{Performance-constrained carbon optimization}

\begin{figure*}
    \centering
    \begin{subfigure}{0.31\textwidth}
        \centering
        \scalebox{0.57}{\includegraphics{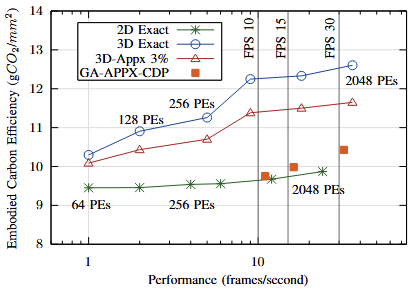}}
        \caption{45nm}
        \label{fig:efficiency_45nm}
    \end{subfigure}
    \hspace{0.02\textwidth}
    \begin{subfigure}{0.31\textwidth}
        \centering
        \scalebox{0.57}{\includegraphics{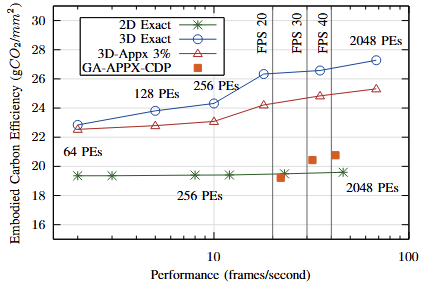}}
        \caption{14nm}
        \label{fig:efficiency_14nm}
    \end{subfigure}
    \hspace{0.025\textwidth}
    \begin{subfigure}{0.31\textwidth}
        \centering
        \scalebox{0.57}{\includegraphics{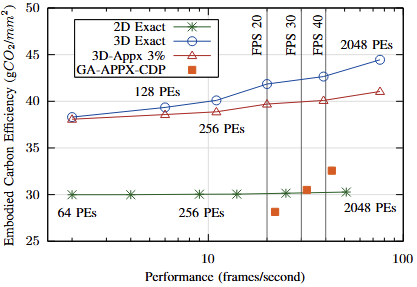}}
        \caption{7nm}
        \label{fig:efficiency_7nm}
    \end{subfigure}
    \caption{Embodied carbon efficiency (g$CO_2$/$mm^2$) and performance (frames/second) for VGG16 across different technology nodes. The x-axis is in logarithmic scale. GA-APPX-CDP is our proposed method, which minimizes CDP while meeting realistic FPS constraints.}
    \label{fig:efficiency}
\end{figure*}

Previous studies~\cite{gupta2022act} have shown that modern DNN accelerators are often overdesigned, delivering more performance than needed for edge applications while significantly increasing their embodied carbon footprint. To analyze the trade-off between carbon efficiency and performance, we modeled NVDLA-like architectures with PE arrays ranging from 64 to 2048 PEs, scaling in powers of two. The local and global buffer sizes were proportionally adjusted based on MAC array dimensions, following NVIDIA's specifications~\cite{nvdla2017}.  

Our evaluation compares four design approaches:
\begin{inparaenum}[(1)]
 \item 2D architectures with exact multipliers (2D Exact); 
 \item 3D architectures with exact multipliers (3D Exact);
 \item 3D architectures with approximate multipliers (3D-Appx), which allow up to 3\% inference accuracy degradation; and
 \item our genetic algorithm-optimized solution (GA-APPX-CDP), which minimizes CDP while meeting realistic FPS constraints. The GA-APPX-CDP optimization ensures that the final design satisfies accuracy drop thresholds and practical edge system requirements. According to the technology node, we applied FPS targets of 10, 15, 20, 30, and 40 to guide our optimization process.
\end{inparaenum}

Figure~\ref{fig:efficiency} shows that conventional 3D accelerators consistently outperform 2D designs in computational performance due to vertical integration and reduced interconnect latency. However, this comes at a significant carbon cost, as 3D designs exhibit higher embodied carbon per unit area. Integrating approximate multipliers reduces this overhead, lowering the carbon footprint of 3D accelerators while maintaining competitive performance. Despite these improvements, the carbon gap between 3D and 2D architectures remains notable, reflecting the manufacturing costs of 3D stacking. Our GA-APPX-CDP solutions achieve a better balance, retaining the performance benefits of 3D integration while approaching the carbon efficiency of 2D designs. This advantage is particularly prevalent in advanced technology nodes, such as 14nm and 7nm, where our approach meets target FPS thresholds of 20, 30, and 40 with significantly better carbon efficiency than conventional 3D implementations. For instance, at 7nm with a 20 FPS target, our solution achieves 32\% better carbon efficiency compared to an exact 3D accelerator with similar performance and 7\% lower carbon per unit area than a 2D architecture satisfying the same FPS threshold.

Due to space limitations, we present results only for VGG16, but the observed trends are consistent across all evaluated DNN models.

\section{Conclusion}

This work shows that approximate computing and genetic algorithm-based design space exploration effectively reduce the embodied carbon footprint of 3D DNN accelerators while preserving high computational efficiency. Our results show that carbon reductions of up to 30\% at 14nm and 25\% at 45nm can be achieved by integrating approximate multipliers, with accuracy degradation up to 3\%. Moreover, at 7nm with a 20 FPS constraint, our optimized designs improve carbon efficiency by 32\% compared to exact 3D implementations and lower carbon per unit area by 7\% relative to 2D designs. These findings confirm that a co-optimized approach, combining approximate computing and CDP-driven optimization, enables sustainable and efficient AI hardware.

\section*{Acknowledgments}

This work has been supported by grant NSF CCF 2324854.

\scriptsize

\end{document}